\documentclass[%
 reprint,
 superscriptaddress,
nofootinbib,
 amsmath,amssymb,
 aps,
prc,
]{revtex4-2}

\usepackage{color} 
\usepackage{graphicx}
\usepackage{dcolumn}
\usepackage{bm}

\newcommand{\Ve}[1]{\ensuremath{\boldsymbol{#1}}}
\newcommand{\be}{\begin{equation}}
\newcommand{\ee}{\end{equation}}
\newcommand{\eq}[1]{\begin{align}#1\end{align}}
\usepackage{braket}
\usepackage[normalem]{ulem}

\begin{document}


\title{Uncovering the sign of nuclear deformations: Determination of prolate or oblate shape via low-energy $\alpha$ inelastic scattering}

\author{Shin Watanabe}
\email[]{s-watanabe@gifu-nct.ac.jp}
\thanks{This work was partly conducted during a sabbatical leave at the Departamento de 
F\'isica At\'omica, Molecular y Nuclear, Universidad de Sevilla.}
\affiliation{National Institute of Technology (KOSEN), Gifu College, Motosu 501-0495, Japan}
\affiliation{RIKEN Nishina Center, Wako 351-0198, Japan}

\author{Yoshiki Suzuki}
\affiliation{Research Center for Nuclear Physics (RCNP), Osaka University, Ibaraki 567-0047, Japan}

\author{Masaaki Kimura}
\affiliation{RIKEN Nishina Center, Wako 351-0198, Japan}

\author{Kazuyuki Ogata}
\affiliation{Department of Physics, Kyushu University, Fukuoka 819-0385, Japan}
\affiliation{Research Center for Nuclear Physics (RCNP), Osaka University, Ibaraki 567-0047, Japan}

\date{\today}

\begin{abstract}
\noindent{\bf Background:}
Understanding nuclear shape is a crucial problem in nuclear physics. In particular, determining the sign
of quadrupole deformation, i.e., whether prolate or oblate,  remains a challenging problem.

\noindent{\bf Purpose:}
Our aim is to propose a method for determining the sign of quadrupole deformation using $\alpha$ inelastic scattering data
and to demonstrate its effectiveness.

\noindent{\bf Method:}
Our approach is the standard coupled-channel method based on the macroscopic model.
We utilize the nuclear reorientation effect, a phenomenon associated with the self-coupling of excited states,
as a probe sensitive to the sign of deformation.

\noindent{\bf Results:}
We first provide an overview of how the reorientation effect influences inelastic scattering cross sections,
and numerically confirm its validity in realistic cases.
We then demonstrate that the sign of deformation can be uniquely determined from inelastic scattering cross section data.

\noindent{\bf Conclusion:}
Our technique offers a systematic approach for determining the sign of deformation in
both stable and unstable nuclei.
The broad applicability of $\alpha$ inelastic scattering will make it a valuable tool to study shape of nuclei, especially unstable nuclei.

\end{abstract}

\maketitle


\section{Introduction}

In nuclear physics, the concept of shape is essential for understanding the complex dynamics of nucleons.
It is well known that most nuclei have non spherical shapes,
reflecting a wide variety of the underlying nucleon dynamics.
The occurrence of nuclear deformation is closely related
to the rotational mode to restore the rotational symmetry (known as the Nambu-Goldstone mode).
This leads to collective rotation and significantly affects the energy levels,
nuclear sizes, and electromagnetic properties~\cite{Bohr75}.

The most essential part of nuclear deformation is quantified by the quadrupole deformation parameter $\beta_2$, if we assume axial symmetry.
Its magnitude $|\beta_2|$ has been deduced mainly through the electric quadrupole transition probabilities
$B(E2)$ between the $0^+$ ground state and the first $2^+$ state~\cite{RAMAN2001}.
Additionally, inelastic scattering experiments, enhanced significantly by the development of the coupled-channel method, have become an important tool to determine $|\beta_2|$~\cite{Tamura65,BROEK1965259,HENDRIE1968127,Alder66,Alder75,Hillis76,BAKER1977135,Mac77,Mac78,Satchler83,Motobayashi95}.
Very recently, total reaction cross sections were also used to deduce $|\beta_2|$.
For example, the systematic study of neutron-rich Ne and Mg isotopes~\cite{TAKECHI2012357,Takechi14,Minomo12,Horiuchi12,Watanabe14}
provided useful information on the boundary of the island of inversion~\cite{Warburton1990} and its extension.
Thus, the systematic determination of $\beta_2$ is an important subject in nuclear physics.

Despite the numerous studies on $|\beta_2|$,
the determination of its sign, whether prolate or oblate, is still a challenging task in nuclear physics.
A nucleus is elongated along its symmetry axis with $\beta_2>0$ (prolate nucleus),
while it is flattened along it with $\beta_2<0$ (oblate nucleus).
It has been theoretically predicted that prolate nuclei dominate oblate ones in the chart of nuclides~\cite{Stoitsov03,Hila07},
but this has not been experimentally confirmed and the underlying mechanism is still unknown.
The difficulty in determining the sign arises because many observables are sensitive to the square of $\beta_2$,
but not to its sign, e.g., the transition probability $B(E2)\propto\beta_2^2$
and the mean radius $\braket{r^2}\approx\braket{r^2}_0(1+\frac{5}{4\pi}\beta_2^2)$, 
where $\braket{r^2}_0$ represents the mean radius in the spherical limit.
The most decisive way of determining the sign is to measure electric quadrupole
 moments ($Q$-moments) through hyperfine spectroscopy.
However, this measurement is difficult to apply to neutron-rich unstable nuclei.
Another way of determining the sign is Coulomb excitation experiment conducted
 at ``safe'' sub-barrier energy~\cite{Alder66,Alder75}.
In the early stage of this method, it was applied to stable target nuclei to extract the quadrupole collectivity.
In the next stage, the Coulomb excitation technique was applied to unstable nuclei
 using the intermediate-energy radioactive isotope beam~\cite{Motobayashi95}.
However, intermediate-energy reactions cannot carry the information on sign.
While the $Q$-moment and Coulomb excitation experiments provide crucial information
 about nuclear deformation including the sign,
their application to unstable nuclei presents significant challenges.

To overcome this difficulty, we propose a method of determining the sign of deformation
by using low-energy $\alpha$ inelastic scattering.
On the basis of the well-established experimental and theoretical techniques~\cite{HENDRIE1968127,BROEK1965259,Tamura65,Satchler83},
we focus on the reorientation effect (RE), which is known as the self coupling of excited states.
It is known that the Coulomb RE plays a critical role in sub-barrier energy
Coulomb excitation processes~\cite{Alder66,Alder75} and fusion reactions~\cite{Hagino97}.
On the other hand, our study actively utilizes the nuclear RE~\cite{Hillis76,BAKER1977135} to determine the sign.
As an example, we consider $\alpha+{}^{154}$Sm scattering at 50~MeV,
for which the experimental data are available~\cite{HENDRIE1968127}.
$^{154}$Sm is often considered to be a prolate-deformed nucleus~\cite{Hila07,Vil78}
whose $Q$-moment for the $2_1^+$ state is $-1.87(4)$~b~\cite{Stone05}.
We will discuss the feasibility of using the RE in $\alpha$ inelastic scattering to determine the sign.

The paper is organized as follows.
In Sec.~\ref{sec:overview_RE}, we provide an overview of the RE in a simplified model.
Section~\ref{sec:DWBS} clarifies the origin of the RE
and discusses its qualitative effect on cross sections in the distorted-wave Born series.
In Sec.~\ref{sec:CC}, we present a standard coupled-channel formalism based on 
the macroscopic model, which will be used in our analysis. We also explain
the origin of the RE in the context of coupled-channel equations
in Secs. \ref{sec:CC} and \ref{sec:TidalSpin}.
The results are presented in Sec.~\ref{sec:results}.
First, in Sec.~\ref{sec:ND}, we provide the numerical evidence that the intuitive behavior
shown in Sec.~\ref{sec:overview_RE} does indeed hold in realistic cases.
Following this, in Sec.~\ref{sec:determining}, we focus on determining the sign of
 deformation using experimental data.
In Sec.~\ref{sec:discussion}, we discuss the feasibility of this method
for determining the sign of deformation for unstable nuclei.
Finally, the paper is concluded in Sec~\ref{sec:summary}.

\section{Overview of the reorientation effect} \label{sec:overview_RE}
We consider $\alpha$ scattering on a deformed target to clarify the RE.
To simplify the discussion, we restrict the model space to the $0^+$ and $2^+$ states
of the axially deformed target nucleus with only the quadrupole deformation $\beta_2$.

\subsection{Distorted wave Born Series} \label{sec:DWBS}
The rationale behind distinguishing the sign of deformation parameter is elucidated as follows.
Let us compare the scattering of $\alpha$ off the prolate ($\beta_2=\beta_+>0$) and oblate ($\beta_2=\beta_-<0$) target nuclei
with the same magnitude of the quadrupole deformation ($\beta_+=|\beta_-|$).
The total deformed potential $V$ can be decomposed into a spherical component $U$
and a residual component $W$, hence $V=U+W$.
Up to the first order of $\beta_2$, $W$ is proportional to $\beta_2$; and the distorted waves $\chi$ are evaluated with $U$
and independent of $\beta_2$.
For the transition from the $i=0^+$ to the $f=2^+$ state, the distorted-wave Born series 
of the $T$-matrix element is given by
\eq{
T_{fi}(\beta_2)=T_{fi}^{(1)}(\beta_2)+T_{fi}^{(2)}(\beta_2^2)+\cdots,\label{eq:Tmat}
}
where
\eq{
T_{fi}^{(1)}(\beta_2)&=\Braket{\chi_{f}^{(-)}\psi_{f}|W|\chi_i^{(+)}\psi_i},\label{eq:T1}\\
T_{fi}^{(2)}(\beta_2^2)&=\Braket{\chi_{f}^{(-)}\psi_{f}|W\tilde{G}^{(+)}W|\chi_i^{(+)}\psi_i}.\label{eq:T2}
}
The superscript $(n)$ denotes the $n$th perturbation,
$\psi_i$ ($\psi_f$) is the initial (final) target state,
$\chi_i^{(+)}$ ($\chi_f^{(-)}$) is the initial outgoing (final incoming) distorted wave,
and $\tilde{G}^{(+)}$ is the distorted-wave Green function~\cite{Satchler83}.
It should be noted that, for $\beta_\pm$, the functions of $\chi$, $\psi$, and $\tilde{G}^{(+)}$ are identical,
whereas the phase of $W$ is opposite.

In Eq.~\eqref{eq:Tmat}, the first term $T_{fi}^{(1)}$ corresponds to the first-order distorted-wave Born approximation, often referred to as simply ``DWBA."
Since $T_{fi}^{(1)}$ is proportional to $\beta_2$,
$T_{fi}^{(1)}$ has an opposite phase for $\beta_\pm$, that is, $T_{fi}^{(1)}(\beta_+)=-T_{fi}^{(1)}(\beta_-)$.
Hence, the DWBA yields the identical cross sections, making it impossible to distinguish between prolate and oblate shapes.
\footnote{When the non central terms are considered in the diagonal potentials of the excited states,
the condition $T_{fi}^{(1)}(\beta_+)=-T_{fi}^{(1)}(\beta_-)$ is not satisfied even in the DWBA. This will be discussed again in Sec.~\ref{sec:CC}.}

This situation changes at the second-order perturbation due to the RE.
Since $T_{fi}^{(2)}$ is proportional to $\beta_2^2$, this 
term is exactly the same for $\beta_\pm$ including the phase.
Eventually, the coherent sum of the first- and second-order $T$-matrix elements will
carry the information of the sign.
$T_{fi}^{(2)}$ acts destructively for $\beta_+$ and constructively for $\beta_-$
if we assume that $\tilde{G}^{(+)}$ does not affect the phase.
This leads to the inequality: $|T_{fi}^{(1)}(\beta_+)+T_{fi}^{(2)}(\beta_+^2)|^2<|T_{fi}^{(1)}(\beta_-)+T_{fi}^{(2)}(\beta_-^2)|^2$,
which will be numerically demonstrated below.
The difference for $\beta_\pm$ may be more noticeable if the higher-order part, i.e., multiple scattering, becomes more important.
We expect that low-energy $\alpha$ inelastic scattering can be a good probe
because the multi step processes may be more important than in Coulomb excitation processes,
which will be shown in Fig.~\ref{fig:xsec_2+_NC}.

\subsection{Coupled-channel formalism} \label{sec:CC}
We recapitulate a standard coupled-channel formalism based on the macroscopic model,
which is used for the numerical demonstration of the above-mentioned idea and for more detailed analysis.
The details of the formulation are given in, for example, Refs.~\cite{Tamura65,Satchler83}.
Dynamics of $\alpha$ scattering on a deformed target is governed by the coupled-channel equation
\eq{
&\left[-\frac{\hbar^2}{2\mu}\frac{d^2}{dR^2}+\frac{\hbar^2}{2\mu}\frac{L(L+1)}{R^2}
+U_{I L,I L}^{(J)}(R)-E_I\right]\chi_{IL}^{(J)}(R)\nonumber\\
&\hspace{5mm}=-\sum_{(I'L')\ne(I L)}U_{I L,I'L'}^{(J)}(R)\chi_{I'L'}^{(J)}(R),\label{eq:CC-eq}
}
where $\mu$ is the reduced mass regarding a projectile and a target,
$I$ and $L$ denote the spin of a target and the angular momentum associated with the relative coordinate ${\Ve R}$
between a projectile and a target,
and $J$ is the total angular momentum of the scattering system.
The wave function $\chi_{IL}^{(J)}$ describes the relative motion in the ($IL$) channel,
$E_I$ is the relative energy with the target in the spin $I$ state,
and $U_{I L,I'L'}^{(J)}$ stands for the coupling potential including nuclear and Coulomb interactions.

The coupling potentials $U_{I L,I'L'}^{(J)}$ can be obtained from the deformed potential
$V(R,\theta)$, where $\theta$ is the direction of $\alpha$ in the body-fixed frame.
To simplify the discussion, we consider the Woods-Saxon potential and
its derivative, which correspond to the first-order approximation of $V(R,\theta)$ for $\beta_2$, as radial form factors:
\eq{
V_\mathrm{N}^{(0)}(R)&=\frac{-(V_0+iW_0)}{1+\exp{\left[\frac{R-R_0}{a}\right]}},\label{eq:V0}\\
V_\mathrm{N}^{(2)}(R)&=-\beta_2R_0\sqrt{\frac{5}{4\pi}}\frac{dV_\mathrm{N}^{(0)}(R)}{dR},\label{eq:V2}
}
where $V_0$ and $W_0$ represent the potential depths for the real and imaginary parts, respectively,
$R_0$ is the radius parameter, and $a$ is the diffuseness parameter.
Similarly, the Coulomb part $V_\mathrm{C}^{(\lambda)}$ is considered in the first-order approximation~\cite{Tamura65}.
Then, we can obtain the coupling potential as
\be
U_{I L,I'L'}^{(J)}(R)
=V^{(0)}(R)\delta_{LL'}\delta_{II'}+ZV^{(2)}(R),\label{eq:Ubar}
\ee
where $V^{(\lambda)}$ ($\lambda=0$ or 2) is the sum of the nuclear and Coulomb parts.
The factor $Z$, which comes from the angular integration, is independent of $\beta_2$.

We clarify the RE in terms of the coupled-channel equation~\eqref{eq:CC-eq}.
In Eq.~\eqref{eq:CC-eq}, the RE is induced by the $2^+$ diagonal potential $U_{2 L,2L}^{(J)}$ (left-hand side)
and the $2^+$ off-diagonal potential $U_{2 L,2L'}^{(J)}$ (right-hand side).
Obviously, the positive and negative signs of $\beta_2$ give the opposite phase of $V^{(2)}$ [Eq.~\eqref{eq:V2}],
affecting both the $2^+$ diagonal and off-diagonal potentials.
It is worth noting that the change in the magnetic quantum number $M_I$ of the $2^+$ excited state
 is described by both of them in the present scheme.
If the RE is neglected, namely $V^{(2)}=0$ for $I=I'=2$ in Eq.~\eqref{eq:Ubar},
the coupled-channel equation becomes symmetric for $\beta_\pm$.
Consequently, $\beta_\pm$ yield exactly the same cross section in the absence of the RE.
Thus, $V^{(2)}$ brings about the difference between $\beta_\pm$ as demonstrated below.

\subsection{Tidal spin} \label{sec:TidalSpin}
To better understand the reorientation effect within the coupled-channel formalism,
the concept of tidal spin~\cite{Gom86} is quite valuable. The tidal spin operator
is defined as the projection of the channel spin operator (target spin $I$ in the present case) along the relative coordinate $\Ve{R}$.
The discussion using the tidal spin $K$ has an advantage when the coupling potential $U_{I L,I'L'}^{(J)}$ [Eq.~\eqref{eq:Ubar}] is strong because $U_{I L,I'L'}^{(J)}$ can be partially diagonalized in the tidal-spin representation.

Using the tidal spin basis and the isocentrifugal approximation~\cite{Gom86},
the coupled-channel equation~\eqref{eq:CC-eq} is reduced to
\eq{
&\left[-\frac{\hbar^2}{2\mu}\frac{d^2}{dR^2}+\frac{\hbar^2}{2\mu}\frac{\bar{L}(\bar{L}+1)}{R^2}
+\bar{U}_{I,I}^{(K)}(R)-E_I\right]\bar{\chi}_{IK}^{(J)}(R)\nonumber\\
&\hspace{5mm}=-\sum_{I'\ne I}\bar{U}_{I,I'}^{(K)}(R)\bar{\chi}_{I'K}^{(J)}(R)\label{eq:CC-eq2}
}
with 
\eq{
\bar{U}^K_{I,I'}(R)
&=\sum_\lambda\frac{\sqrt{4\pi}}{\hat{I}\hat{\lambda}}
\braket{I'K\lambda 0|IK} \nonumber\\
&\hspace{10mm}\times\braket{I||Y_\lambda ||I'}V^{(\lambda)}(R),
}
where $\bar{L}$ is the average angular momentum, $\hat{x}=\sqrt{2x+1}$,
and $\bar{U}_{I,I}^{(K)}$ and $\bar{\chi}_{IK}^{(J)}$ denote the coupling potential 
and the relative wave function in the tidal spin representation, respectively.
It should be noted that $\bar{L}$ and $K$ are conserved in Eq.~\eqref{eq:CC-eq2}.

For the specific case of the $0^+$ to $2^+$ transition, Eq.~\eqref{eq:CC-eq2} formally corresponds to 
a coupled-channel scattering problem involving spinless states,
where only $K = 0$ is relevant.
Under this assumption, the reorientation term of the $2^+$ states becomes an additional
central term for the excited state, described as
\eq{
\bar{U}^0_{0,0}(R)&=V^{(0)}(R),\\
\bar{U}^0_{2,2}(R)&=V^{(0)}(R)+\frac{2}{7}V^{(2)}(R).\label{eq:Ubar2}
}
Equation~\eqref{eq:Ubar2} allows us to understand the sensitivity of the calculations to the sign of
the quadrupole deformation parameter $\beta_2$.
When $\beta_2>0$, the second term adds an attractive contribution (real part)
and increases absorption (imaginary part).
On the other hand, when $\beta_2<0$,
the second term introduces a repulsive contribution (real part)
and reduces absorption (imaginary part). 
Consequently, these differences lead to observable changes in the cross sections.

\section{Results}\label{sec:results}
In order to numerically demonstrate the validity of the discussion in Sec.~\ref{sec:overview_RE},
$\alpha$ scattering on a deformed $^{154}$Sm target at 50~MeV is considered as an example.
To simplify the discussion, we restrict the model space to the $0^+$ and $2^+$ states of the target nucleus,
including only the quadrupole deformation $\beta_2$, as was done in Sec.~\ref{sec:overview_RE}.
The Woods-Saxon potential parameters $V_0=65.9$ MeV, $W_0=27.3$ MeV, $R_0=1.44\times154^{1/3}$ fm, and $a=0.70$ fm
are used for all the calculations below.
These parameters are determined from the elastic scattering cross section data~\cite{HENDRIE1968127} in Sec.~\ref{sec:determining}.

\subsection{Numerical demonstration of reorientation effect}\label{sec:ND}
In this subsection, we assume that $U_{0L,0L}^{(J)}=U_{2L,2L}^{(J)}=V^{(0)}$
so that the condition $T_{fi}^{(1)}(\beta_+)=-T_{fi}^{(1)}(\beta_-)$ is satisfied in the DWBA calculation. This assumption highlights the non trivial part of the RE, as shown below.

Figure~\ref{fig:xsec_2+_1step2step} illustrates the results of one-step and two-step calculations
for the scattering of $\alpha+{}^{154}$Sm at 50 MeV.
For simplicity, only the nuclear excitation is considered in this subsection.
The dashed line represents the one-step calculation (DWBA), corresponding to the first term of Eq.~\eqref{eq:Tmat}.
No difference is observed for $\beta_\pm$ in this calculation.
However, they differ in the two-step calculations, which correspond
 to the sum of the first and second terms of Eq.~\eqref{eq:Tmat}.
As discussed in the context of Eq.~\eqref{eq:Tmat}, the second term acts destructively for $\beta_+$
and constructively for $\beta_-$, leading to a distinct difference from forward to backward angles
as shown by the solid ($\beta_+$) and dashed ($\beta_-$) lines, respectively.
This second term is the primary contributor to the RE.

\begin{figure}[htbp]
\includegraphics[width=0.45\textwidth,clip]{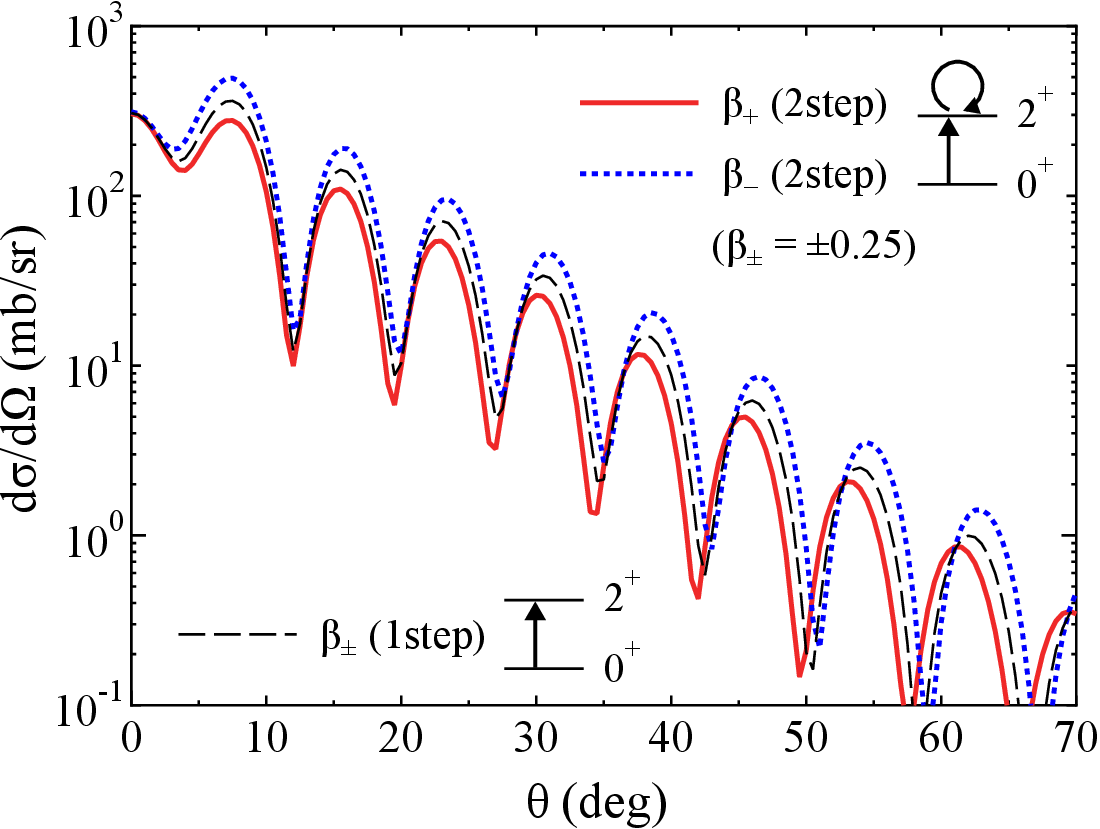}
\caption{One- and two-step calculations of the inelastic scattering cross section for the scattering of $\alpha+{}^{154}$Sm at 50 MeV.
The dashed line correspond to the one-step calculation (DWBA), showing no distinction between $\beta_\pm=\pm0.25$.
The solid and dashed lines represent the two-step calculations for $\beta_+$ and $\beta_-$, respectively.
}
\label{fig:xsec_2+_1step2step}
\end{figure}

\begin{figure}[htbp]
\includegraphics[width=0.45\textwidth,clip]{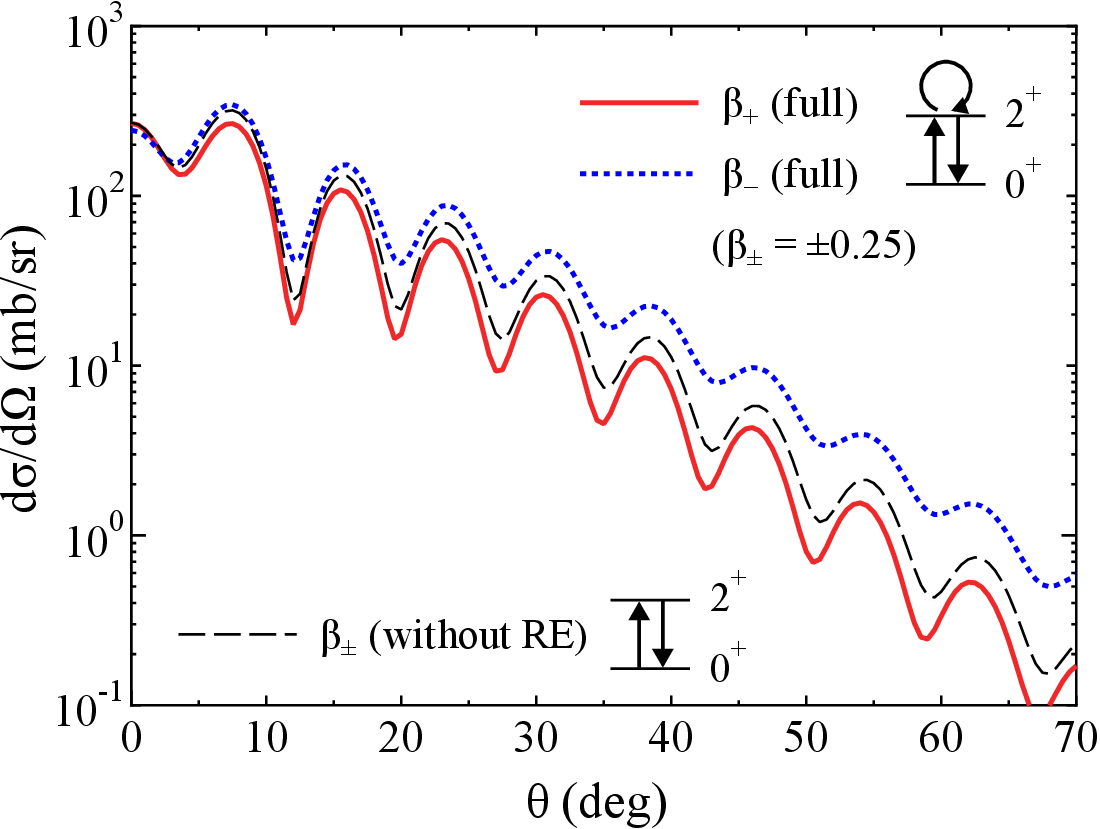}
\caption{Same as Fig.~\ref{fig:xsec_2+_1step2step}, but the channel-coupling (CC) effect is fully taken into account.
The solid and dotted lines represent the full CC calculation for $\beta_+=+0.25$ and  $\beta_-=-0.25$, respectively.
The dashed line denotes the calculation for $\beta_\pm$ without the RE.
}
\label{fig:xsec_2+_woRE_full}
\end{figure}

We further investigate the impact of the higher-order terms in Eq.~\eqref{eq:Tmat}.
Figure~\ref{fig:xsec_2+_woRE_full} presents the inelastic scattering cross section calculated by incorporating
the channel coupling in all orders.
The solid and dashed lines indicate the full calculations for $\beta_+$ and $\beta_-$, respectively.
Even after considering the channel-coupling effect, the discussion in the previous paragraph is still valid;
the RE acts destructively for $\beta_+$ and constructively for $\beta_-$, leading to a distinct difference.
The channel-coupling effects merely smoothen the cross section from the one- or two-step calculations
 in Fig.~\ref{fig:xsec_2+_1step2step}.
If the RE is neglected, $\beta_\pm$ yield exactly the same cross section as shown by the dashed line.
In the absence of RE, even-order perturbations vanish,
and only odd-order perturbations are summed coherently in Eq.~\eqref{eq:Tmat}.
Eventually, the resultant $T$-matrix elements for $\beta_\pm$ are identical except for their opposite phases.
However, with the RE, even-order terms interfere differently for $\beta_\pm$,
enabling us to determine the sign of $\beta_2$.

\subsection{Determining the sign of deformation from experimental data}\label{sec:determining}
In this subsection, we demonstrate how we can determine the sign of deformation from experimental data.
Non-central terms of $U_{2L,2L}^{(J)}$ [Eq.~\eqref{eq:Ubar}] are fully taken into account in the following analysis.

\begin{figure}[htbp]
\includegraphics[width=0.45\textwidth,clip]{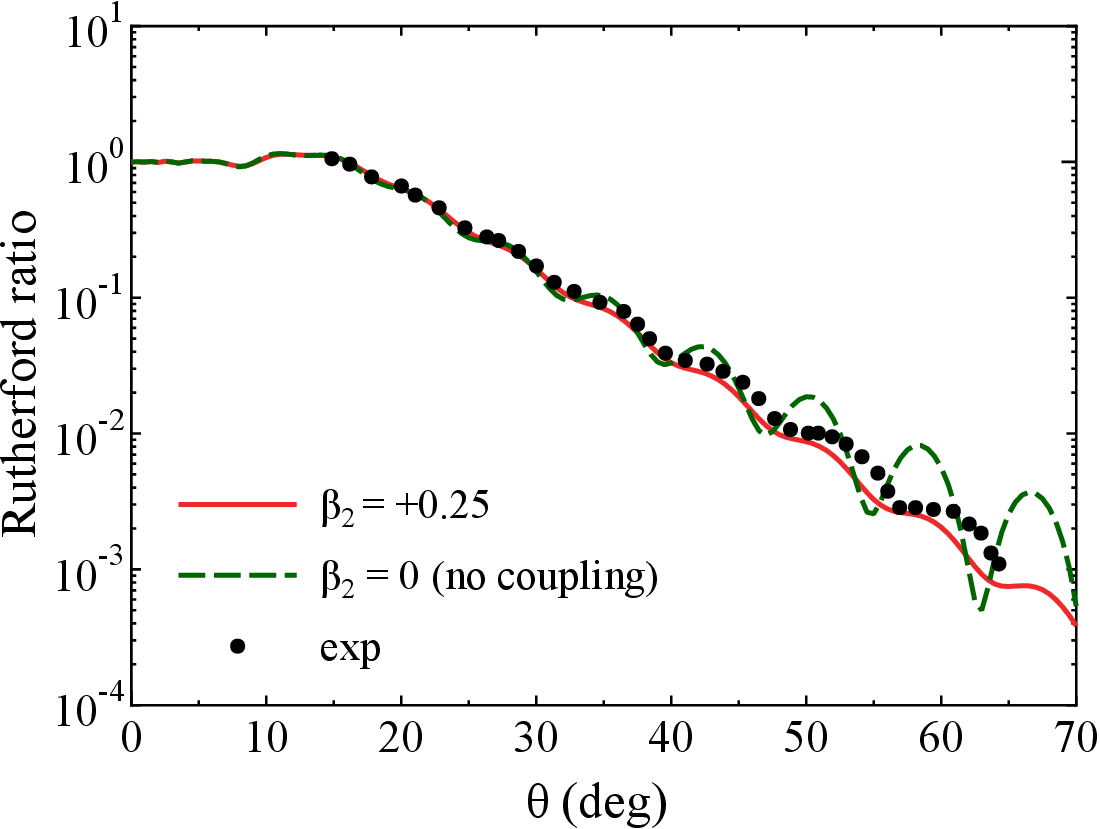}
\caption{Elastic scattering cross section for the scattering of $\alpha+{}^{154}$Sm at 50 MeV.
The solid and dashed lines represent the result with $\beta_2= +0.25$
and $\beta_2=0$ (no coupling), respectively.
The experimental data are taken from Ref.~\cite{HENDRIE1968127}.}
\label{fig:xsec_el}
\end{figure}

First, we fix the potential parameters ($V_0$, $W_0$, $R_0$, $a$) by the fitting to the elastic scattering cross section.
Figure~\ref{fig:xsec_el} presents the elastic scattering cross section for $\alpha+{}^{154}$Sm scattering at 50 MeV.
We adopt the potential parameters from Ref.~\cite{HENDRIE1968127} with a slight modification; a larger
 diffuseness parameter of 0.7 fm is used to effectively account
 for the coupling effect of higher spin states on the monopole potential.
 The dashed line represents the result with $\beta_2=0$ (no coupling),
 reproducing the experimental data~\cite{HENDRIE1968127} except for the strong oscillation at backward angles.
 This oscillation diminishes when the $2^+$ coupling is included, as shown by the solid line.

\begin{figure}[htbp]
\includegraphics[width=0.45\textwidth,clip]{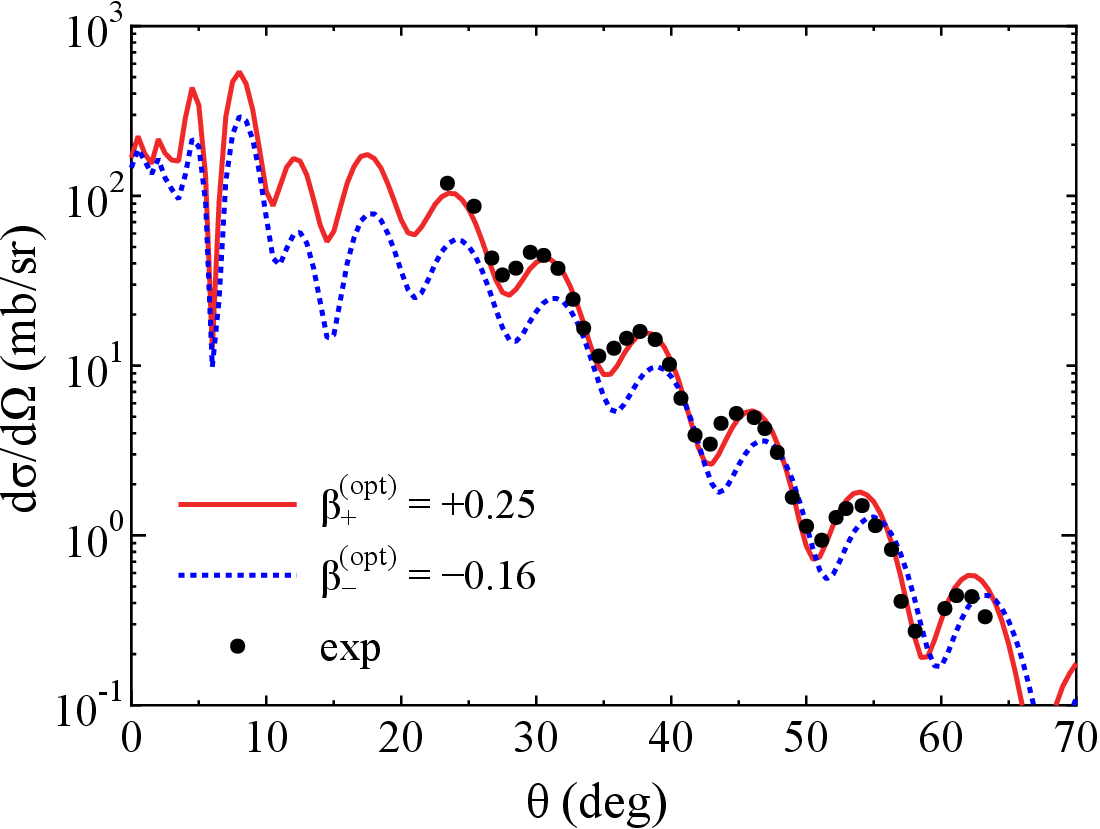}
\caption{Inelastic scattering cross section for the scattering of $\alpha+{}^{154}$Sm at 50 MeV.
The solid and dotted lines represent the optimized results for $\beta_+^\mathrm{(opt)}=+0.25$
and $\beta_-^\mathrm{(opt)}=-0.16$, respectively.
The experimental data are taken from Ref.~\cite{HENDRIE1968127}.}
\label{fig:xsec_2+}
\end{figure}

Next, we vary $\beta_2$ from negative to positive values to determine the optimized deformation parameters
for the experimental data of the inelastic scattering ($\beta_-^\mathrm{(opt)}$ and $\beta_+^\mathrm{(opt)}$)~\cite{HENDRIE1968127}.
Note that $^{154}$Sm has a prolate shape ($\beta_2>0$)~\cite{Stone05}.
Figure~\ref{fig:xsec_2+} illustrates the angular distribution of the inelastic scattering cross section for
the $\alpha+{}^{154}$Sm scattering to the $2_1^+$ state.
The solid line represents the best fitted result with $\beta_+^\mathrm{(opt)}=+0.25$,
while the dotted line corresponds to $\beta_-^\mathrm{(opt)}=-0.16$, which is the best fit among the negative sign results of $\beta_2$.
The result with $\beta_+^\mathrm{(opt)}$ is in good agreement with the experimental data
 from the forward to backward angles.
In contrast, the result with $\beta_-^\mathrm{(opt)}$ deviates from the data even at the forward angle.
This difference is due to the RE for $\beta_\pm^\mathrm{(opt)}$.
As discussed in Sec.~\ref{sec:ND}, the cross section becomes larger
for oblate deformation than the prolate one when $|\beta_2|$ is the same.
Therefore, the smaller magnitude of $|\beta_-^\mathrm{(opt)}|=0.16$ is necessary
to reproduce the experimental data overall.
However, it fails to reproduce the rise of the cross section at forward angle $\theta\approx25^\circ$
and the position of the diffraction minimum (the effective reaction radius).
Thus, the sign of deformation can be determined
by analyzing the RE in low-energy $\alpha$ inelastic scattering.

\begin{figure}[htbp]
\includegraphics[width=0.45\textwidth,clip]{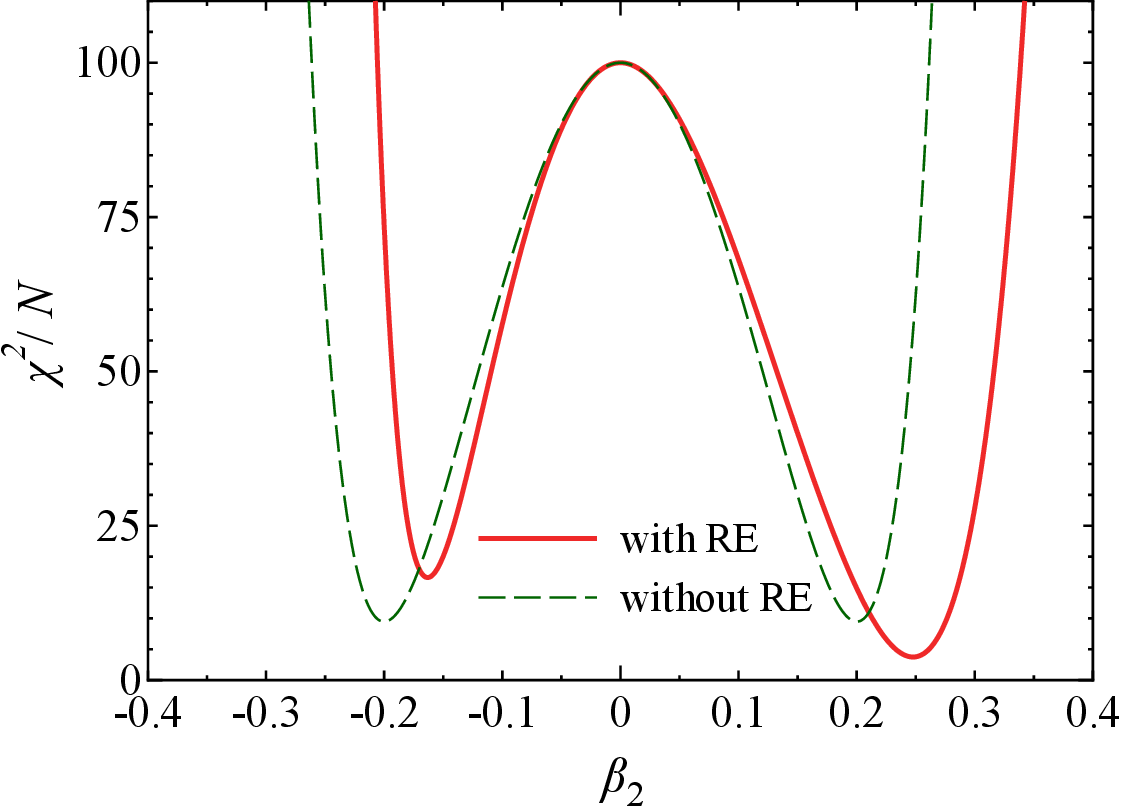}
\caption{$\chi^2/N$ as a function of the deformation parameter $\beta_2$. The solid and dashed lines correspond
to the results with and without the RE, respectively.}
\label{fig:bet-chi2}
\end{figure}

To quantify the deviation of the calculation from the experimental data, chi-squared ($\chi^2$) is defined as
\be
\chi^2(\beta_2)=\sum_{i=1}^{N}\left(\frac{\sigma_i^\mathrm{exp}-\sigma_i^\mathrm{theor}(\beta_2)}{\Delta_i^\mathrm{exp}}\right)^2,
\ee
where $N$ is the number of the data points, $\sigma_i^\mathrm{exp}$ and $\sigma_i^\mathrm{theor}$ are
the experimental and theoretical cross sections, respectively,
and $\Delta_i^\mathrm{exp}$ represents the error of the experimental data; 10~\% error is assumed to calculate $\chi^2$.
Figure~\ref{fig:bet-chi2} shows the $\chi^2$ result
 where the solid and dashed lines represent the results with and without the RE, respectively.
It is evident that the same absolute value of $|\beta^{(\pm)}|$ does not lead to the equivalent $\chi^2$ value with the RE.
Consequently, the positively optimized deformation parameter ($\beta_+^\mathrm{(opt)}=+0.25$) yields a smaller $\chi^2$
than the negatively optimized one ($\beta_-^\mathrm{(opt)}=-0.16$).
Thus, the sign of the deformation is determined to be positive
through the comparison between the theoretical result and the inelastic scattering cross section data.

To further refine the analysis, we treated the diffuseness parameter $a$ as an adjustable parameter,
independently fitting it to both elastic and inelastic scattering data for each value of $\beta_\pm^\mathrm{(opt)}$. This ensures that the diffuseness parameter 
$a$ is optimized in an unbiased manner for both positive and negative values of $\beta_\pm^\mathrm{(opt)}$. It is interesting that $a=0.70$ fm is an optimal value not only
for $\beta_+^\mathrm{(opt)}$ but also for $\beta_-^\mathrm{(opt)}$, because larger values of $a$
rapidly increase $\chi^2$ for the elastic scattering cross section.
By using different parameter sets (e.g., $\beta_2=-0.20$ and $a=0.76$ fm),
it is possible to reproduce the inelastic scattering cross section with negative values of $\beta_2$,
but it fails to reproduce the elastic scattering cross section (absolute value and phase) simultaneously. These findings further support the reliability of the sing determined
in our method.

In Table~\ref{tbl:beta}, the extracted deformation parameters $\beta_\pm^\mathrm{(opt)}$
are compared with the experimental data deduced from \mbox{$B(E2)$$\uparrow$}~\cite{RAMAN2001}.
The experimental nuclear deformation parameter $\bar{\beta}_\mathrm{exp}=0.339$ is transformed into the potential deformation parameter
by $\beta_\mathrm{exp}=\frac{1.2}{1.44}\bar{\beta}_\mathrm{exp}=0.283$ to make it comparable
with $\beta_\pm^\mathrm{(opt)}$~\cite{HENDRIE1968127}.
It is clear that $\beta_+^\mathrm{(opt)}=0.25$ is much closer to $\beta_\mathrm{exp}=0.283$ than $|\beta_-^\mathrm{(opt)}|=0.16$.
The discrepancy can be understood by referring to Fig.~\ref{fig:bet-chi2}.
The minimum positions of $\beta_2=\pm0.20$ in the dashed line shift towards the positive direction for both $\beta_2$
due to the destructive ($\beta_2>0$) and constructive ($\beta_2<0$) interference of the RE.
This result emphasizes the importance of considering the RE 
to extract the sign and magnitude of the deformation parameter from experiments.
In conclusion, the RE in low-energy $\alpha$ inelastic scattering provides
 an effective means of determining the sign of deformation.

\begin{table}[htbp]
\caption{The comparison between the potential deformation parameters obtained by our analysis
and the experimental data deduced from \mbox{$B(E2)$$\uparrow$} $=4.32$ $e^2\mathrm{b}^2$~\cite{RAMAN2001}. The experimental nuclear deformation parameter
$\bar{\beta}_\mathrm{exp}=0.339$ is transformed into the potential
deformation parameter by $\beta_\mathrm{exp}=\frac{1.2}{1.44}\bar{\beta}_\mathrm{exp}=0.283$~\cite{HENDRIE1968127}.
}
\label{tbl:beta}
\begin{center}
\begin{tabular*}{50mm}{@{\extracolsep{\fill}}ccc}
\hline
$\beta_\mathrm{exp}$ & $\beta_+^\mathrm{(opt)}$ & $|\beta_-^\mathrm{(opt)}|$ \\
\hline
 0.283 & 0.25 & $0.16$ \\
\hline
\end{tabular*}
\end{center}
\end{table}

\section{Discussion}\label{sec:discussion}

\begin{figure}[htbp]
\includegraphics[width=0.45\textwidth,clip]{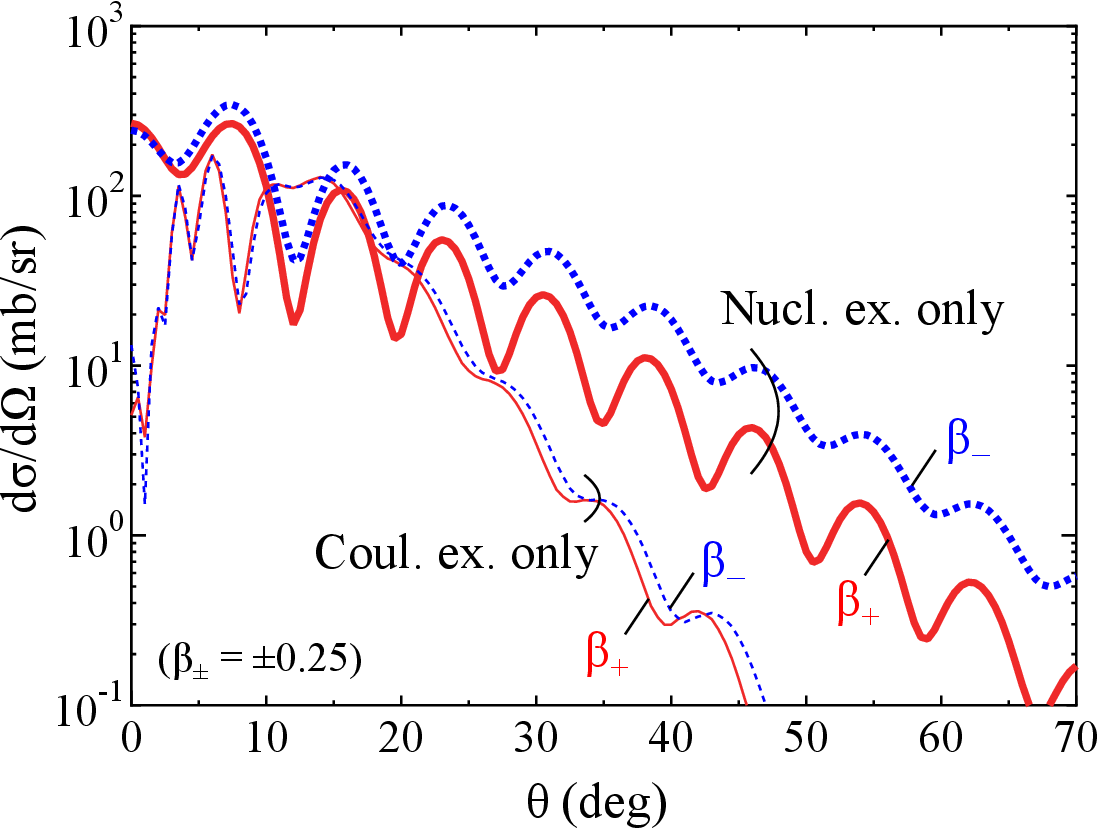}
\caption{
Contributions of nuclear and Coulomb excitations to the inelastic scattering cross section for the scattering of $\alpha+{}^{154}$Sm at 50 MeV.
The thick lines represent calculations with nuclear excitation (Nucl. ex.) only, 
and the thin lines correspond to Coulomb excitation (Coul. ex.) only. 
In both cases, the solid lines indicate $\beta_+=+0.25$ while the dotted lines represent $\beta_-=-0.25$.
}
\label{fig:xsec_2+_NC}
\end{figure}

The distinct difference in inelastic scattering cross sections for \(\beta_\pm\) is due to the nuclear RE rather than the Coulomb RE.
This is clearly demonstrated in Fig.~\ref{fig:xsec_2+_NC}, where the contributions from nuclear and Coulomb excitations are decomposed.
The calculations with only nuclear excitation (Nucl. ex.) are indicated by the thick lines,
 whereas those involving only Coulomb excitation (Coul. ex.) are shown by the thin lines.
In each case, the solid line denotes $\beta_+=+0.25$ and the dotted line corresponds to $\beta_-=-0.25$.
This separation highlights that the overall pattern is primarily shaped by nuclear excitation,
while Coulomb excitation uniformly enhances the cross sections at forward angles, where the Coulomb RE is negligible.
These findings suggest that $\alpha$ inelastic scattering could be an effective tool
 for detecting the sign of deformation through nuclear RE.
Furthermore, a lower incident energy, such as $E_\alpha=50$ MeV, is suitable for observing these effects,
as the distinction of $\beta_\pm$ in cross sections becomes less pronounced at higher energies
 where higher-order effects are less important.

\begin{figure}[htbp]
\includegraphics[width=0.45\textwidth,clip]{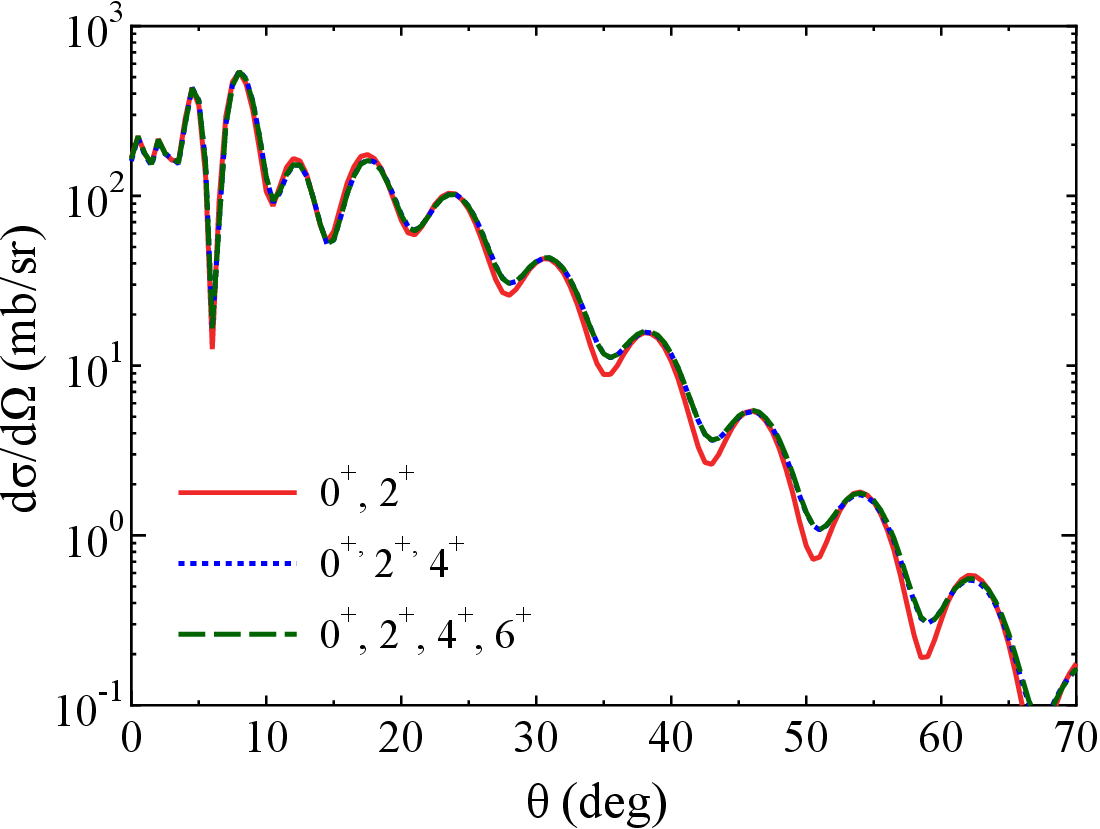}
\caption{Model space dependence on the inelastic scattering cross section for $\alpha+{}^{154}$Sm scattering at 50 MeV.
The solid line represents the $0^+$ and $2^+$ states,
the dotted line includes states up to the $4^+$ state, and the dashed line includes states up to the $6^+$ state.
The deformation parameter $\beta_+=+0.25$ is taken.
}
\label{fig:xsec_2+_0246_beta+}
\end{figure}

In Fig.~\ref{fig:xsec_2+_0246_beta+}, the model space dependence is illustrated.
Various model spaces are indicated by the different line styles: the solid line for ($0^+$, $2^+$),
the dotted line for ($0^+$, $2^+$, $4^+$), and the dashed line for ($0^+$, $2^+$, $4^+$, $6^+$) models.
The results indicate that expanding the model space has a minimal impact on the final results.
This finding confirms the robustness of our previous discussions
and supports the consistency of the discussion in more realistic models.

\begin{figure}[htbp]
\includegraphics[width=0.45\textwidth,clip]{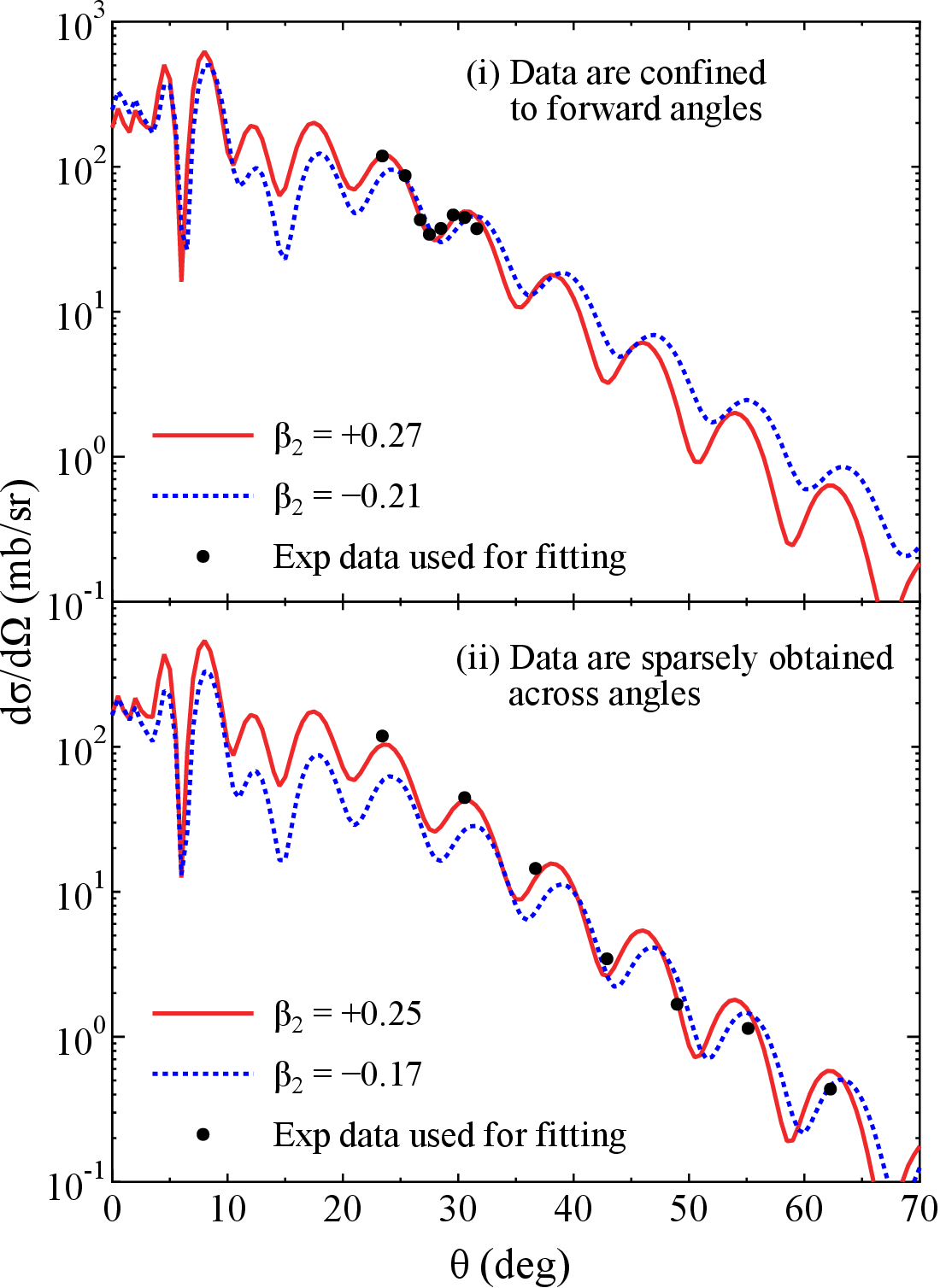}
\caption{The $\chi^2$ fitting under two assumptions for $\alpha+{}^{154}$Sm scattering at 50 MeV:
 (i) Data are confined to forward angles and
 (ii) data are sparsely obtained across angles.
 For each assumption, the solid and dotted lines correspond to the
 positively optimized and negatively optimized results, respectively.
}
\label{fig:xsec_2+_thtmax32_every6}
\end{figure}

Finally, we discuss the feasibility of this method for determining the sign of deformation for unstable nuclei
in inverse kinematics experiments.
In these experiments, we often face challenges, particularly
\begin{itemize}
 \item[(i)] data are confined to forward angles;
 \item[(ii)] data are sparsely obtained across angles
\end{itemize}
In Fig.~\ref{fig:xsec_2+_thtmax32_every6},
the top panel illustrates challenge (i),
where the deformation parameters $\beta_\pm$ are optimized for forward angles, which cover the two peaks.
The optimized deformation parameters are slightly modified compared to those in Fig.~\ref{fig:xsec_2+}:
$\beta_+^{(\mathrm{opt})}=+0.27$ and $\beta_-^{(\mathrm{opt})}=-0.21$ for challenge (i), and 
$\beta_+^{(\mathrm{opt})}=+0.25$ and $\beta_-^{(\mathrm{opt})}=-0.17$ for challenge (ii).
There is still a difference between the prolate and oblate results,
 as the effective reaction radius is different: a smaller $|\beta_2|$ leads to a smaller reaction radius for oblate nucleus.
However, the cross sections at backward angles will help us
to reliably determine the sign of deformation for unstable nuclei even though they are sparsely obtained.
Challenge (ii) is depicted in the bottom panel, where the data are selected at every sixth interval from the original ones.
Though the data are limited, there remains a clear distinction between the prolate and oblate results.
This discrepancy arises from the RE altering the overall slope of the cross section.
The difference in slope is also shown at the backward angle in panel (i),
 where the experimental data are unavailable under this assumption.
Therefore, we recommend extending measurements to backward angles, say, transfer momentum
 $q\approx2.5$ fm${}^{-1}$ ($\theta_\mathrm{max}\approx50^\circ$ in the present system),
even in cases of sparse data collection.
Such an approach would likely yield more definitive results
in determining the sign of deformation for unstable nuclei.

\section{Summary}\label{sec:summary}
In summary, we have proposed a new method for determining the sign of deformation
by analyzing the reorientation effect in low-energy $\alpha$ inelastic scattering. 
We elucidate the rationale for distinguishing the sign of deformation
using the distorted-wave Born series and numerically confirm
that the effect does indeed hold in realistic cases.
We apply this method to $\alpha+{}^{154}$Sm scattering at 50 MeV
to demonstrate its practical effectiveness.
The results show a clear distinction between the inelastic scattering cross sections optimized
for positive and negative deformations owing to the reorientation effect.
We conclude that the reorientation effect in low-energy $\alpha$ inelastic scattering
is a decisive factor in determining the sign of deformation.
We expect that systematic measurements of $\alpha$ inelastic scattering cross sections will advance
the systematic determination of signs of deformation, serving as a reliable indicator of shell evolution.

\begin{acknowledgments}
We would like to thank  N. Aoi, E. Ideguchi, S. Shimoura, and T. Motobayashi for valuable discussions.
One of the authors, S.W., thanks A. M. Moro and the faculty and staff
at Universidad de Sevilla for their hospitality during
his sabbatical stay, which enabled the completion of this work.
This work was supported by Japan Society for the Promotion of
Science (JSPS) KAKENHI Grants No. JP22K14043 and No. JP20K03971.
\end{acknowledgments}

\bibliography{./ref}

\end{document}